# Hydrosomes: Optically trapped femtoliter containers for single molecule studies and microfluidics

J. E. Reiner[1], A. M. Crawford[1], R. B. Kishore[1], Lori. S. Goldner[1], M. K. Gilson[2], K. Helmerson[1]

[1]Physics Laboratory, National Institute of Standards and Technology, 100 Bureau Dr., Gaithersburg, MD 20899

[2]Center for Advanced Research in Biotechnology, 9600 Gudelsky Dr. Rockville, MD 20850

We demonstrate a novel technique for creating, manipulating, and combining femtoliter volume chemical containers. Possible uses include creating controlled chemical reactions involving small quantities of reagent, and studying the dynamics of single molecules within the containers by fluorescence imaging techniques. The containers, which we call hydrosomes, are surfactant stabilized aqueous droplets in a low index-of-refraction fluorocarbon medium. The index of refraction mismatch between the container and fluorocarbon is such that individual hydrosomes can be optically trapped by single focus laser beams, i.e. optical tweezers. Here we trap and manipulate individual hydrosomes. We demonstrate a controlled chemical reaction by the fusion of a hydrosome containing DNA segments approximately 1000 base pairs in length with a hydrosome containing YOYO-1, a DNA intercalating dye. We furthermore detect the fluorescence from single dye molecules in a hydrosome, and observe single pair fluorescence resonance energy transfer (spFRET) from Cy3-Cy5 molecules attached to a single-stranded 16mer DNA molecule.

## Introduction

One of the most important functions of a cell is to act as a small container for sub-cellular structures. Increasingly, researchers have found ways to incorporate structures that hold small volumes into fabricated devices. Small containers holding femtoliter volumes have great potential utility in a wide variety of chemical and biological applications.[1,2]

In the analytical chemistry community, there is a need for techniques to rapidly and efficiently process samples that involve chemical reactions. Rapid processing requires both the expeditious handling of the samples and that the chemical reactions take place quickly. Researchers seek to work with ever-smaller volumes since both physical movement and diffusional mixing of chemicals in very small volumes can be quite rapid. Efficient processing is necessary for cost effectiveness and applications where only a small amount of analyte may be available.

In biological applications, small containers can provide a means of confining single molecules inside of a femtoliter measurement volume typical of a confocal microscope.[3,4] Techniques for optically observing single molecules are changing and extending our understanding of molecular processes in biology.[5] Although measurements can be made on single molecules in a dilute solution as they diffuse through the measurement volume, this type of measurement cannot follow the dynamics of a single molecule for longer than the transit time through the measurement volume, usually on the order of a few milliseconds.[6] To study dynamics on longer time scales, methods have been developed to immobilize and isolate or confine the molecule or molecular complex under study while still permitting physiological interactions and optical observations to occur. Current immobilization strategies include binding[7-9] or adsorption[10-13] of molecules on a surface; immobilization in porous materials,[4, 14] and more recently; encapsulation in a surface-tethered vesicle.[3,15,16] The first two cases introduce the likelihood of artifacts because of molecular interactions with an environment (the surface) that is both inhomogeneous and difficult to characterize; this shortcoming has been the single largest impediment to and criticism of single molecule techniques. The third case, which is physiologically more relevant, will be discussed in more detail below.

One approach to working with small volumes focuses on the controlled flow of liquids through microchannels in glass, plastic and other solid substances.[17] This approach is comparatively well developed and has met with significant successes. However, it requires the fabrication of a dedicated device for each type of analysis and can lead to troublesome surface interactions due to the large surface-to-volume ratios of the elongated channels. Volumes in microchannels are usually not small enough for diffusional mixing to be very efficient unless specialized structures such as 'chaotic mixers' are incorporated into the design.[18]

A second, complementary, approach to working with small volumes involves the use of small containers. Depending on the application, nano-containers may be used in conjunction with or in place of microfluidic channels. In general, a miniature container for holding pico- to femtoliter volumes of liquid should satisfy three main requirements. (1) The container should be closed or sufficiently isolated from the environment that the substances held in the container do not escape into the surrounding medium, by evaporation or diffusion. (2) It must be possible to access the contents of the container in order to add reagents as required by a given experimental protocol. (3) The contents of individual containers should be independently controllable so that distinct reactions can take place in separate containers.

Several different types of nanocontainers have been explored to date. Perhaps the most straightforward approach has been to fabricate nanovials in solid substrates, in analogy to microtitre plates.[19] These are open containers with volumes down to the hundreds of femtoliters. Recently this type of container has been used for detection of single molecules even at a relatively high concentration of analyte.[20] It is difficult for them to satisfy requirements (1) and (2) however as, except in special circumstances, molecules are free to diffuse in and out of the vial.

Another type of nanocontainer is suggested by the observation that living cells routinely satisfy the three criteria listed above. Several groups are investigating the use of giant (cell sized) liposomes as nanofluidic containers.[21-23] Liposomes are closed structures consisting of a phospholipid bilayer membrane, either uni- or multi-lamellar, that isolate an aqueous interior from an aqueous external environment. Several groups have started encapsulating single molecules inside liposomes and tethering the liposomes to a surface for study with a confocal system.[3,15,16] This technique has been successful in localizing single molecules within the confocal volume and shielding the molecule from surface interactions that have troubled direct tethering techniques. Thus far no interaction between the lipid membrane and the encapsulated molecules has been observed, but significant challenges to the use of liposomes as nano-containers remain.[3] One involves the formation of uniformly sized liposomes quickly and on demand; recent work with flow focusing in microfluidic channels has started to address this issue.[24] It is also difficult to incorporate reagents inside the liposomes in known quantities and, washing away excess reagent adds an extra step to the sample preparation that can lead to loss of sample. Once a liposome is formed, it is difficult to gain further access to its contents for introducing new reagents. Merging two liposomes together involves disrupting the membrane, which may cause leakage of the contents.

An emulsion in which one liquid forms drops in another immiscible liquid is a simple way of forming very small closed containers. Studies have demonstrated transcription and translation of single genes[25] and single molecule PCR[26] within the droplets of bulk emulsions. However, it is often desirable to control both the movement and the contents of individual drops in the emulsion independently. Several steps have been made in this direction. Rotman[27] used aqueous drops stuck to a coverslip in oil to look at fluorescent signals from single enzymes. He et al.[28] have selectively encapsulated single cells and sub-cellular structures into water droplets in oil created with microfluidic channels and then trapped the cells inside the water drops with optical tweezers. Water droplets in oil have been manipulated by pushing with optical tweezers[29] and dielectrophoresis in microfluidic channels.[30] Single dye molecules inside a water droplet levitated in air by an electrodynamic trap were detected by Barnes et al.[31] However, the rapid evaporation of the droplets in air is problematic for measurements of biological molecules.

The present study describes a technique for independently controlling drops in an emulsion. Here, micron-sized aqueous droplets, termed "hydrosomes," are immersed in an inert, immiscible, non-aqueous medium. Each hydrosome can contain its own distinct chemical contents. Optical tweezers are used to trap and drag droplets around and, as shown in the results section, can be used to cause two droplets to merge and thus mix their contents. Spectroscopic methods and fluorescent techniques can be used to analyze the contents of an optically trapped hydrosome. This method satisfies the three main requirements put forth for small containers and is relatively easy to implement.

## Experimental Section

### Materials

For our non-aqueous medium we use fluorinert FC-77 which is a fluorocarbon liquid purchased from 3M Co., St. Paul, MN. Triton X-100, Phosphate buffer solution (PBS), Sulforhodamine-B (SRB) and Ethylenediaminetetaacetic acid (EDTA) were purchased from Sigma Chemical Co., St. Louis, MO. YOYO-1 iodide was purchased from Molecular Probes, Eugene, OR. Tris-HCL was purchased from Gibco BRL, Life Technologies Inc, Gaithersburg, MD. Log DNA ladder was purchased from New England Biolabs, Beverly, MA. Single-stranded 16mer DNA with Cy3 attached to the 5' end and Cy5 attached to the 3' end was purchased from Sigma-Aldrich, Woodlands, TX. β-mercaptoethanol was purchased from Pierce, Rockford, IL. Glucose oxidase and catalase were purchased from Roche Diagnostics Co., Indianapolis, IN. Red fluorescent protein (rDs Red2 protein) was purchased from BD Biosciences Clontech, San Diego, CA.

Several different aqueous solutions were prepared for this work. Red fluorescent protein (RFP) was mixed with PBS, 10 mmol/L EDTA, and 10 % glycerol at pH=7.45. 10 mg/L DNA solution and 4 µmol/L YOYO-1 iodide solution were prepared in 10 mmol/L Hepes buffer at pH 7.4. SRB and the fluorescently labeled ssDNA were diluted in a TRIS buffer containing an oxygen scavenging system. The oxygen scavenging system consisted of 4 % beta-mercaptoethanol, 50 µg/mL glucose oxidase, 10 µg/mL catalase, 18 % (w:w) glucose in a 10 mmol/L TRIS buffer with 150 mmol/L NaCl and 2 mmol/L EDTA at pH 7.5.

**Preparation of Hydrosomes**

Each of these solutions was mixed with Triton X-100 at 0.1 % by volume. FC-77 was filtered through a 0.2 µm filter. 100 µL of aqueous solution was mixed with 1 mL of filtered FC-77. Sonicating this mixture in an ultrasonic bath (5510, Branson, Danbury, CT) for 30 seconds formed hydrosomes less than 2 µm in diameter. Sonication was performed at room temperature for solutions containing SRB, YOYO-1 or DNA and in an ice bath for RFP. The resulting hydrosomes were stable at room temperature, as expected given that the solubility of water in the fluorocarbon is at the level of a few ppm. However, the droplets were observed to shrink over a period of several minutes when held in the optical trap, presumably due to weak absorption of the trapping light, which results in heating. This shrinkage was markedly reduced or even eliminated by mixing a surfactant, such as Tween-20 or Triton X-100, into the water.

**Optical Manipulation and Detection of Hydrosomes**

We use a single focus laser trap, or optical tweezers, to trap and remotely manipulate the hydrosomes. Optical tweezers rely on the increased polarizability of the object to be trapped compared to the surrounding medium, such that the energy of interaction between the object and the laser field is a minimum. That is, the object to be trapped must have an index of refraction higher than the surrounding medium. In the case of our hydrosomes, the index of refraction of water is 1.33 compared to 1.29 for FC-77. Therefore the hydrosomes are easily manipulated with optical tweezers.

A schematic of the optical setup is shown in Figure 1. The trapping apparatus is based on an inverted microscope (Axiovert S100, Zeiss, Germany). We use two independent optical tweezers for the fusion demonstration. The traps are generated by an infrared laser with $\lambda = 1.064$ µm and a maximum output power of 5 watts (YLM Series, IPG Photonics, Oxford, MA). A dichroic mirror (XF2017, Omega Optical, Brattleboro, VT) reflects about 500 mW per trap into the back aperture of a 100x, 1.4 numerical aperture, oil-immersion objective lens (Plan-Apochromat, Zeiss). One trap is fixed and the other can be swept, via a movable mirror (M1 in figure 1), across the field of view, so that two hydrosomes can be trapped simultaneously and moved with respect to each other. Fluorescence excitation is provided with either a fiber-coupled tunable argon-ion laser (43 Series Ion Laser, Melles Griot, Carlsbad, CA) or a diode-pumped solid-state (DPSS) green laser with a wavelength of 532 nm and an output power of 120 mW (HPM-120, Extreme Lasers, Seabrook, TX). A charge coupled device (CCD) camera (XC-

ST50, Sony, Japan) was used for white light video microscopy and an intensified CCD camera (I-Pentamax, Roper Scientific, Trenton, NJ) for fluorescence imaging.

Single molecule fluorescence detection is performed with a modified confocal setup. One of the optical trapping beams is overlapped with the green laser, which is defocused with respect to the trapping laser to a size of approximately 5 μm in order to approximate uniform illumination of a 1 μm diameter spot at the focus of the trapping laser. Light collected by the objective is focused onto a 100 μm pinhole just outside the microscope. The light from the pinhole is filtered (XB11, Omega Optical and HQ550LP, Chroma Technology, Rockingham, VT) to allow only the fluorescent light of interest to be focused down onto the active area (diameter = 175 μm) of an avalanche photodiode (APD) (SPCM-AQR-14, Perkin Elmer, Wellesley, MA). In the case of spFRET detection, the 100 μm pinhole is removed in order to eliminate problems associated with chromatic aberration and mirror M3 is replaced with a dichroic mirror (XF2021, Omega Optical) that reflects donor fluorescence from Cy3 molecules and transmits acceptor fluorescence from Cy5 molecules. A second APD with a fluorescent filter (3RD650LP, Omega Optical) in front. Both APDs send TTL pulses to a counting card (NCI-6602, National Instruments, Austin, TX) controlled by software (Labview 6.1, National Instruments). Photons are counted in 1 ms time bins.

**Fluorescent Detection of Binding within a Hydrosome**

Hydrosomes, when mixed in bulk, tend to fuse spontaneously. This makes it difficult to perform a controlled mixing reaction with two different populations of hydrosomes made from ultrasonic agitation. When we made two separate solutions of hydrosomes and mixed them together on the same slide we found that the majority of hydrosomes larger than 1 μm had been made up of a collection of fused hydrosomes. We performed a modified version of a mixing experiment by using the photobleaching properties of a DNA intercalating dye.

A 3 μm diameter hydrosome containing both DNA and dye was prepared by randomly fusing a collection of ≈1 micron diameter hydrosomes, which contained either DNA ladder segments ≈ 1 kbp in length or YOYO-1 intercalating dye. The concentration of the DNA ladder segments in solution before forming the hydrosomes was 10 mg/L, which implies that a 1 μm diameter hydrosome contains approximately 5 DNA ladder molecules or a total of about 5,000 base pairs. The concentration of YOYO-1 in solution prior to forming the hydrosomes was 4 μmol/L, which implies that a 1 μm diameter hydrosome contains approximately 1000 dye molecules. We verified that the individual hydrosomes, either the ones containing only DNA or only dye, did not fluoresce when excited by 488 nm light. The 3 μm hydrosome did fluoresce and hence it contained both DNA and intercalated dye, which was expected since it was prepared from a random selection of hydrosomes. Based on the random selection we expect that the 3 μm hydrosome contained 70 ± 17 DNA molecules. In the binding experiment, one of these hydrosomes was fully photobleached, and it was then fused with a smaller hydrosome containing only unbleached YOYO-1 dye.

**Single-Molecule Detection in a Hydrosome**

A microscope slide containing a collection of hydrosomes under bright field illumination is translated until the optical tweezers trap a single hydrosome. The bright field illumination is then blocked and the optical path to the APD is opened. After 2-4 seconds a shutter is opened causing the trapped hydrosome to be illuminated by the green laser. Fluorescence is collected until all the dye molecules in the hydrosome have photobleached, typically 5 s - 20 s. The hydrosome is then released from the trap and the process can be repeated with a fresh hydrosome.

# Results

**Controlled Fusion of Hydrosomes with Mixing and a Binding Reaction**

To evaluate the applicability of hydrosomes as ultra-small volume containers for chemistry, we performed a controlled reaction by fusing two hydrosomes and thus allowing their individual contents to mix and react. The reaction demonstrated was the binding of an intercalating dye to DNA. Figure 2 is a sequence of video images showing the fusion of two hydrosomes held in independent optical tweezers. Fusion of hydrosomes differs from that of liposomes in two major ways. First, hydrosomes fuse spontaneously when they are brought into contact[32], unlike liposomes where fusion has to be induced by an external device such as a microelectrode[20] or a pulsed laser beam.[21] Second, if molecules do not partition into the FC-77 (i.e. the hydrosomes do not leak) then there will be no loss of encapsulated reagent during the fusion process.

Figure 3 shows a binding reaction initiated by the fusion of two hydrosomes containing different reagents. The 3 μm diameter hydrosome containing both DNA and YOYO-1 intercalating dye was held in an optical trap and photobleached by exposure to 488 nm light. The fluorescence was monitored and observed to extinguish below a detectable level in approximately 15 seconds due to photobleaching. Subsequent excitation, even if the hydrosome was first kept in the dark for up to 30 minutes, produced no detectable fluorescence, indicating persistent photobleaching of the dye present in this hydrosome. Next, ≈1 μm diameter hydrosomes, containing only YOYO-1 dye, were transported under flow at a velocity of 10 μm/s through the region containing the 3 μm hydrosome. The trapped 3 μm hydrosome was maneuvered to intersect the path of a dye containing hydrosome (shown by the arrow in figure 3a). When the two hydrosomes nearly overlapped, the optical trap pulled the dye-containing hydrosome into contact with the 3 μm hydrosome and the two hydrosomes subsequently fused. The composite hydrosome was held in the optical trap for 2 minutes before re-exposure to the 488 nm light, at which time the fluorescence of the merged hydrosome was observed. Figure 3b is a fluorescence image of the field containing the two hydrosomes prior to fusion, and Figure 3c shows the same field 2 minutes after fusion. Clearly, significant fluorescence is observed only after fusion of the hydrosomes. We observed that a minimum of 30 seconds was required after fusion before any fluorescence could be detected. Since the mixing time is much shorter than this, the delay presumably is associated with the time required for the fresh dye to replace the photobleached-intercalated dye. After the fluorescence of the merged hydrosome had extinguished due to continued excitation, the controlled reaction could be repeated by fusing another hydrosome containing only dye.

**Fluorescence Detection of Single Molecules in Hydrosomes**

Figure 4 shows three typical measurements of the fluorescence of single sulforhodamine B (SRB) dye molecules contained in hydrosomes. The first few seconds show dark counts from the detector (about 100 counts/s) and the initial jump in intensity occurs when the shutter on the green laser is opened, producing a combination of fluorescence emission and laser background counts. The counts decrease in discrete steps until only dark counts plus background from the excitation light is measured. The discrete drops in emission are indicative of photobleaching of individual dye molecules, and the number of dye molecules in the hydrosome can be counted by the number of photobleaching steps.[3] Thus, Figures 4a, b and c show results from hydrosomes with one, two and three dye molecules, respectively. Results similar to those in figure 4 were obtained for free Cy3 dye, Cy3 attached to a DNA strand, and TMR attached to a DNA strand.

**Detection of Red Fluorescent Protein**

To demonstrate that more fragile biomolecules remain functional during the hydrosome formation and trapping process, we incorporated red fluorescent protein (RFP) into hydrosomes and measured the fluorescence with a confocal detection setup. If the RFP is not damaged during the sonication process then the fluorescence intensity from a solution of RFP and from a large hydrosome that overfills the green laser focal spot should be equal.

In order to eliminate effects that may decrease the detected fluorescence intensity from a large hydrosome we performed a control experiment with a 100 µmol/L solution of SRB. There should be agreement between the observed fluorescence of the bulk solution and large trapped hydrosome. The center of the hydrosome was typically held about 2-3 microns above the glass coverslip. At this height, the detected fluorescence from the large hydrosome was about 50% of the measured intensity in bulk solution. The index of refraction mismatch between the FC-77 and hydrosome may explain this discrepancy because as the hydrosome was brought closer to the cover slip, the detected fluorescence increased until agreement with the bulk solution was finally found when the hydrosome wet to the glass surface.

This experiment was then carried out with a 360 nmol/L concentration of RFP. After correcting for the 50% loss due to the hydrosome-FC-77 index mismatch, we found a lower bound on the survival rate of RFP from ultrasonic agitation to be slightly more than 50%.

**Fluorescence Detection of Single Molecule FRET in Hydrosomes**

We demonstrate the feasibility of single molecule FRET studies within hydrosomes by encapsulating single stranded 16mer DNA molecules with Cy3 attached to the 5' end and Cy5 attached to the 3' end. The green laser excites the FRET pair which results in quenching of the Cy3 donor dye and fluorescence from the Cy5 acceptor dye. Figure 5 illustrates an example of a single pair FRET trajectory.

**Discussion**

This paper demonstrates the creation, individual manipulation, and application of hydrosomes, femtoliter aqueous containers in an immiscible liquid. This approach has a number of strengths. First, in contrast to encapsulation in liposomes, incorporating reagents or single molecules into hydrosomes on their initial formation is as simple as diluting the desired quantities in the buffer which will form the droplets. Second, hydrosomes fuse spontaneously upon contact so mixing different reagents or changing the environment of a single molecule can be performed quickly and easily. Again, this contrasts with liposomes, which require an additional energy input to accomplish fusion, and which are also prone to leakage during the fusion process. Third, hydrosomes are readily trapped and manipulated with optical tweezers, due to the difference between the refractive index of their aqueous contents and the supporting medium, here a fluorocarbon liquid. Fourth, hydrosomes are well suited to study with optical techniques and, of particular interest, can be used to localize single molecules in the diffraction limited volume of a confocal microscope for study by fluorescence techniques.

It is expected that hydrosomes will be useful in a wide range of applications. The facility with which hydrosomes fuse, allows a molecule to be observed as its environment is changed in a controlled way. For example, one might imagine watching a protein unfold as denaturant is added in single-hydrosome aliquots; or bringing receptors and small numbers of ligands together to observe single ligand-receptor binding reactions. Hydrosomes can also be used as a basis for adaptive microfluidics in three dimensions. Thus, whereas most microfluidic devices restrict flows to fixed channels inscribed in a solid substrate, the movement of a hydrosome can be controlled by optical tweezers. As a consequence, fluid transfers can be changed on-the-fly depending upon results obtained during a microfluidic procedure, and there is no requirement that flows be confined to a two-dimensional surface.

The combination of optical trapping with fluorescence microscopy is of particular interest. Early efforts to use these techniques in concert were plagued by technical difficulties due to the increased background from the trapping and position-detection lasers. Two groups[33, 34] have recently demonstrated that optical trapping is compatible with measurements of single-molecule fluorescence as long as appropriate filters are used. These groups focused on using the tweezers for force measurements on a fluorescently tagged molecule or bead tethered to a surface. Here we have shown that the use of optical tweezers is also compatible with coincident and simultaneous detection of untethered single dye molecules held in place by the trapped hydrosome. A possible concern with this approach arises from the observation of Van Dijk *et al.*[33] that some dyes, particularly ones based on carbocyanine, bleach faster when the infrared trapping light is added to the excitation light. We can circumvent this by alternating trapping and excitation light at a sufficiently high repetition rate that measurements of the salient dynamics are unaffected and the hydrosome remains trapped. This can be achieved with the use of a chopper wheel, electro-optic or acousto-optic modulators.

The fluorocarbon suspension medium used here was selected because of its chemical inertness, immiscibility with water, and its refractive index, which is significantly less than that of water and thus enables laser trapping by a single focus laser beam. On the other hand, oxygen is highly soluble in fluorocarbons,[35] so this medium may enhance photo-bleaching. The use of an appropriate oxygen scavenging system, such as the one described in the methods section slows down photobleaching to a time constant of several seconds, which is long enough to measure some single molecule dynamics. We expect the photobleaching lifetime can be further extended with careful refinement of the oxygen scavenging system and a lowering of the fluorescence excitation intensity.

Another concern is that molecules inside the hydrosomes could partition into the water-fluorinert interface or transfer completely into the fluorinert. However, this is not expected to be a problem for hydrophilic molecules. Confocal imaging of a 10 μm diameter hydrosome containing 100 μmol/L SRB showed uniform fluorescence throughout the droplet over a period of several minutes (data not shown). Fluorescence polarization measurements could be performed to confirm that single dye molecules are not localized to the interface. It is worth noting that similar measurements on proteins encapsulated in 100 nm liposomes[3] have shown the proteins do not spend a significant amount of time attached to the bilayer membrane.

Reliable kinetics studies rely on accurate measurements of volumes and concentrations. The uncertainty in measuring the size of hydrosomes is an obstacle that must be overcome in order to perform mixing of different concentrations of reagent. One could imagine incorporating a position detection laser into the current setup that could accurately measure droplet size.[36] Recent work by Link *et al.* demonstrates creation of monodisperse water droplets using microfluidic channels.[37] These droplets are too large for use in single molecule detection and other groups have successfully created drops of only a few microns in diameter with micropipettes and piezoelectric actuators.[38, 39] Micron-sized droplet creation is an active area of research and it seems likely that new ways of creating very small droplets to suite many different types of applications will be available in the future.

**Summary**

We have demonstrated a new technology for creating and individually manipulating femtoliter aqueous containers, which we call hydrosomes. Incorporating reagents or single molecules inside of the hydrosomes is also as simple as diluting the desired quantities in buffer. The hydrosomes fuse spontaneously upon contact so mixing of different reagents, or a change in the environment of a single molecule can be performed quickly and easily. We demonstrated that hydrosomes could be used for creating controlled chemical reactions as well as for localizing single molecules in the diffraction limited volume of a confocal microscope for study by fluorescence techniques. This technique should be useful in a wide range of applications.

## Acknowledgements

We gratefully acknowledge many helpful discussions with Peter Yim, Xiaoyi Zhang, and Thomas Perkins. We also thank Thomas Perkins for suggesting the use of fluorescent proteins in the hydrosomes. This work was supported by the Office of Naval Research and the National Institute of Standards and Technology. Commercial names of materials and apparatus are identified only to specify the experimental procedures. This does not imply a recommendation by NIST, nor does it imply that they are the best available for the purpose.

**Figures**

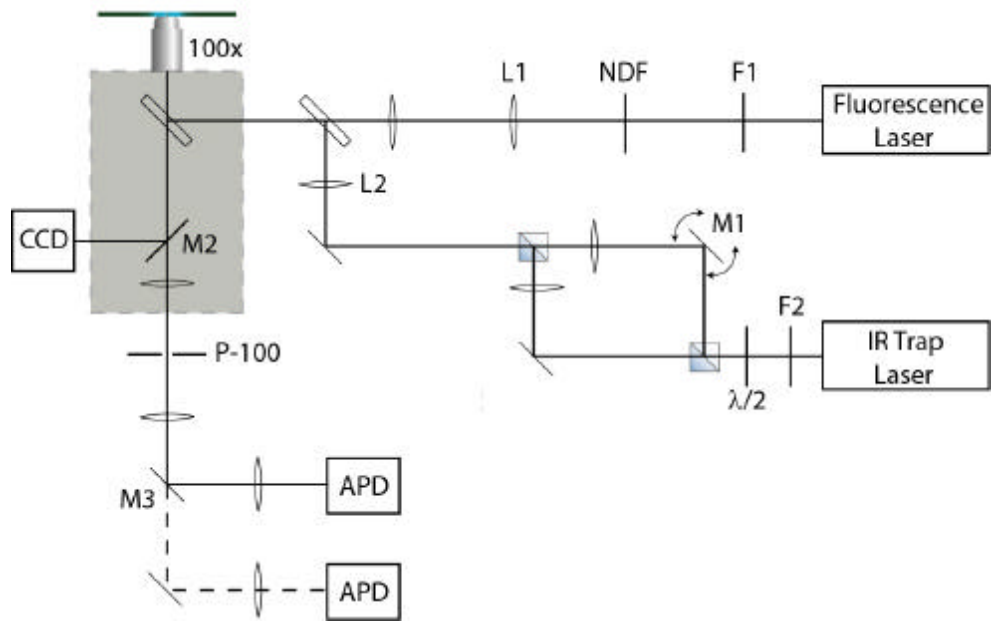

Figure 1. A schematic diagram of the optical setup. F1 and F2 are bandpass filters for 532 nm and 1064 nm light respectively. NDF is a series of neutral density filters. L1 is the first lens of the fluorescence excitation telescope. This lens is present for single molecule confocal detection and removed for wide-field fluorescence detection with the CCD camera. L2 is the second lens of the IR trap beam telescope that is mounted on a 3-axis translation stage for precision overlap of the trap and fluorescence beams. M1 is the movable mirror for the moving optical trap in the fusion experiments. All elements in the shaded grey region are internal to the microscope and M2 is a manual flipper mirror that directs light to the CCD camera or is removed to allow passage to the avalanche photodiode (APD). P-100 is a 100 µm pinhole that is part of the confocal microscope setup. FRET measurements were made by replacing the M3 mirror with a dichroic mirror and a second APD was inserted to capture light from the acceptor fluorophore. Also, the P-100 pinhole was removed for the FRET measurement to avoid issues with chromatic aberrations.

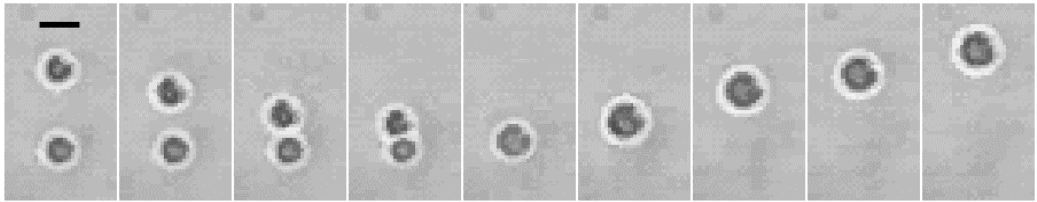

Figure 2. Sequence of video images showing the fusion of two hydrosomes, initially held in independent optical tweezers. The upper hydrosome is translated by the mobile trap to the location of the hydrosome held by the fixed trap, at which point the two hydrosomes fuse into one. The fixed trap is then turned off and the single hydrosome is translated upwards by the mobile trap. The mobile trap (upper) is slightly defocused from the fixed trap (lower). The solid bar in the first picture is 1 µm in length.

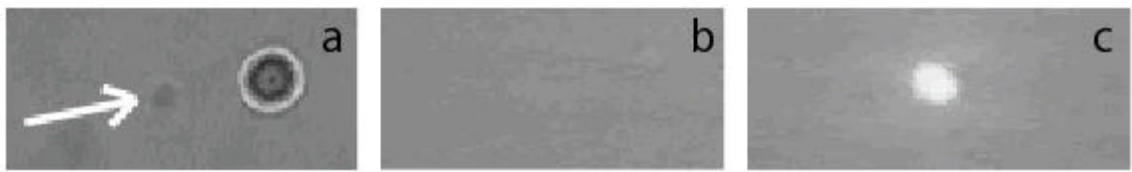

Figure 3. Binding of fluorescent dye, YOYO-1, to DNA in a hydrosome. A: brightfield image of the initial 3 µm diameter hydrosome containing both DNA and photobleached intercalated dye, and the ~1 µm diameter dye-containing hydrosome (arrow) flowing towards it. B: same picture as A under fluorescence excitation. No fluorescence is observed after photobleaching of the 3 µm hydrosome and before fusion. C: fluorescence is observed 2 minutes after fusion of the two hydrosomes.

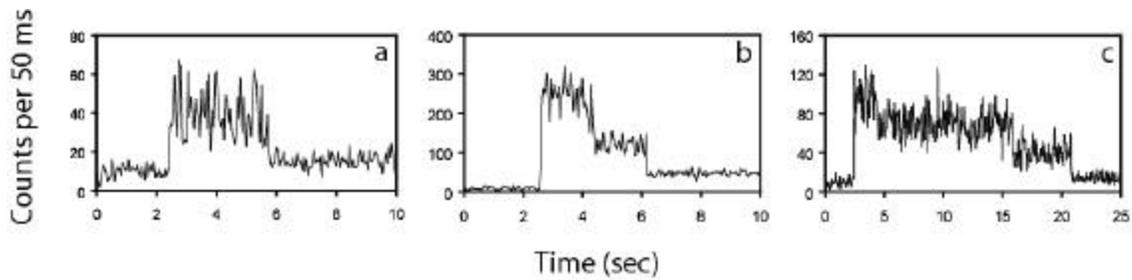

Figure 4. Three examples of single molecule detection in trapped hydrosomes. a, b, and c illustrate the trapping and detection of 1, 2, and 3 sulforhodamine B (SRB) molecules respectively. The measurements are taken with different excitation strengths. For (a) and (c) the laser power sent into the back aperture of the microscope objective was 600 µW and for (b) the power used was 2.5 mW. The different laser powers resulted in different step sizes for photobleaching events. For these measurements a solution of 5 nM SRB was used for which a 1 µm hydrosome is calculated to contain an average of 1.6 dye molecules.

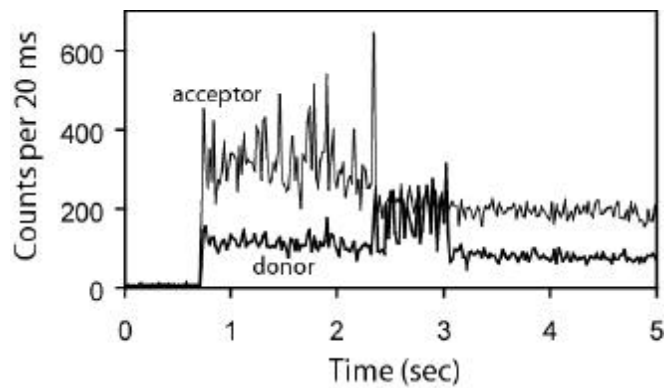

Figure 5. Single molecule FRET data. Single stranded 16mer DNA with a Cy3 molecule attached to the 5' end and a Cy5 molecule attached to the 3' end. The Cy3 donor molecule is quenched by the Cy5 acceptor molecule until photobleaching of the Cy5 molecule occurs at t = 2.3 seconds. The donor molecule fluoresces until t = 3.0 seconds when the donor molecule photobleaches.